\documentclass[conference]{IEEEtran}
\IEEEoverridecommandlockouts
\usepackage{cite}
\usepackage{amsmath,amssymb,amsfonts}
\usepackage{graphicx}
\usepackage{textcomp}
\usepackage{xcolor}
\usepackage{placeins}
\usepackage{algpseudocode}
\usepackage{algorithm} 
\def\BibTeX{{\rm B\kern-.05em{\sc i\kern-.025em b}\kern-.08em
    T\kern-.1667em\lower.7ex\hbox{E}\kern-.125emX}}
\begin{document}

\title{A Gaussian Integral Filter with Multivariate Laplace Process Noise\\
\thanks{This material is based upon work supported by the Air Force Office of Scientific Research under award number FA9550-19-1-0404. Any opinions, finding, and conclusions or recommendations expressed in this material are those of the author(s) and do not necessarily reflect the views of the United States Air Force.}
}

\author{\IEEEauthorblockN{Enrico M. Zucchelli}
\IEEEauthorblockA{\textit{Aerospace Engineering and Engineering Mechanics} \\
\textit{The University of Texas at Austin}\\
Austin, TX, U.S.A \\
enricomarco@utexas.edu}
\and
\IEEEauthorblockN{Brandon A. Jones}
\IEEEauthorblockA{\textit{Aerospace Engineering and Engineering Mechanics} \\
\textit{The University of Texas at Austin}\\
Austin, TX, U.S.A \\
brandon.jones@utexas.edu}
}

\maketitle

\begin{abstract}
This paper introduces the concept of the Gaussian integral filter~(GIF), the limit of the Gaussian sum filter~(GSF) for when the number of mixands tends to infinity. The GIF is obtained via a combination of GSF, quadrature, and interpolation. While it is a very general concept, in this paper the GIF is used to represent multiviariate Laplace (ML) distributions defining the process noise when tracking a maneuvering target. The filter is first applied to a linear three-dimensional toy problem, and then to a maneuvering target tracking problem in Earth orbit. For the more complex maneuvering target tracking problem, the filter requires only 1.4~times the computational resources of an unscented Kalman filter (UKF), while having errors up to 11 times smaller. For the same problem, the UKF slowly diverges.
\end{abstract}


\section{Introduction}
Maneuvering target tracking is a challenging problem that has been widely researched for several decades~\cite{li_2003,li2001survey,li2001bsurvey,li2002survey}. Common approaches include equivalent process noise~\cite{efe1998maneuvering}, adaptive-noise methods~\cite{gholson1977maneuvering}, variable dimension estimators~\cite{bar1982variable,goff2015orbit}, and interacting multiple model (IMM) filters~\cite{li_2005,zucchelli2020tracking}. Most of the above mentioned methods either require fine tuning of parameters, or they adapt to the measurements, causing the approach to be non-Bayesian. A Bayesian method with an explicit transitional prior has the advantage that it can be directly implemented in a multi-target tracking filter such as the probability hypothesis density~(PHD) filter~\cite{clark2006gm} or the generalized labeled multi-Bernoulli~(GLMB) filter~\cite{vo2016efficient,yun2022generalized}. In a Bayesian framework it is often convenient to use heavy-tailed distributions, such as the multivariate Laplace (ML) distribution or Student's $t$-distribution, to represent the maneuvers distribution~\cite{roth2013student,huang2016robust}. Heavy-tailed distributions are more responsive than Gaussian distributions to sudden, large maneuvers, and are thus more robust.
An ML distribution can be described by a continuous Gaussian mixture model (CGMM), which is an infinite sum of Gaussian components; specifically, the ML distribution can be represented by a Gaussian Scale Mixture~(GSM)~\cite{boris2008scale}, which is a subclass of the CGMM.

A Gaussian Sum Filter~(GSF)~\cite{gsf} is a bank of Gaussian filters working in parallel to reproduce non Gaussian distributions more faithfully than a single Gaussian filter would. Depending on the problem, GSFs may be preferred to particle filters (PFs) because they are not subject to sample impoverishment and particle depletion.
In this paper the Gaussian integral filter (GIF) is introduced, which is the limit of the GSF for when the number of components, or mixands, tends to infinity. The result is a combination of GSF with quadrature and interpolation methods over the mixands of the distribution. The GIF is a Bayesian filter that employs a CGMM representation for the prior of the state, for the process noise, for the measurement noise, or for a combination of those distributions.
While the GIF is very generic, and may be used, for example, as an alternative to Gaussian mixture splitting, this paper focuses on how it can be applied to a problem where the process noise is distributed according to an ML distributions.

Huang et al.~\cite{huang2017robust} exploit the GSM formulation of the ML for the process noise to design a Kalman filter based on variational Bayesian methods. The filter is applied to a maneuvering target tracking problem. Wang et al.~\cite{wang2021novel} exploit the same concept, but use the ML distribution for the measurement noise instead; the resulting filter is robust to problems where the measurements have large outliers. Both filters are limited to linear systems, are iterative, and make simplifying assumptions; in addition, they provide Gaussian posterior distributions.

There are three main contributions in this paper. First, the GIF is introduced, a filter that uses a CGMM as prior, process noise, and/or measurement noise, by a combination of GSF, interpolation, and quadrature. To the best of the authors' knowledge, there has been no direct use of a CGMM-based filter to date. Second, the ML-GIF, a GIF that employs the description of the ML as a CGMM for process noise, is described. The only approximations made are the interpolation, the quadrature, and the fact that every single mixand is kept Gaussian during propagation and update. 
Third, the ML-GIF is applied to a challenging maneuvering target tracking problem. The proposed filter requires approximately only 1.3 times the computational time of a UKF when using quadrature and interpolation methods. The method provides a non-Gaussian, possibly heavy-tailed (depending on observability) posterior distribution.

\section{The Continuous Gaussian Mixture Model}
A finite GMM is defined as follows:
\begin{equation}
\label{eq:GMMdef}
    p(x) = \sum_{i=1}^N w_i\, \mathcal{N}\left(x;\mu_i,P_i\right),\qquad \sum_{i=1}^N w_i = 1,
\end{equation}
where $N$ is the number of mixands, $w_i$ is the weight of the $i$\textsuperscript{th} mixand, $\mu_i$ and $P_i$ are the corresponding mean and covariance, respectively.
The CGMM consists of the limit of Eq.~\eqref{eq:GMMdef} when $N$ tends to infinity. For this to be properly defined, a parameterization is required:
\begin{equation}
    \label{eq:CGMMdef}
    p(x) = \int_a^b  \mathcal{N}\left(x;\mu(z),P(z)\right)\,p_z(z) \,dz,
\end{equation}
where $a$ and $b$ are the boundaries of the integral, $z$ is the parameterization variable, and $p_z(z)$ is the probability density function (p.d.f.) of $z$.
To numerically evaluate a CGMM, discretization is needed, which leads to a p.d.f. represented as a finite sum of Gaussian distributions like in~\eqref{eq:GMMdef}. However, at any time, an approximation to the original integral can be recovered by interpolation. It is thus possible to adaptively change the interpolation nodes and achieve arbitrary precision, as well as to sample from the original distribution.

\section{Symmetric ML Distribution as a CGMM}
An ML distribution with mean $\boldsymbol{\mu}\in\mathbb{R}^d$ and variance $\Sigma\in\mathbb{R}^{d\times d}$ has the following p.d.f.:
\begin{align}
    p(\boldsymbol{x}) &= \frac{2}{\sqrt{|2\pi\Sigma|}}\left(\frac{\left(\boldsymbol{x}-\boldsymbol{\mu}\right)^T\Sigma^{-1}\left(\boldsymbol{x}-\boldsymbol{\mu}\right)}{2}\right)^{v/2}\\
   \nonumber &\times\,K_v\left(\sqrt{2\,\left(\boldsymbol{x}-\boldsymbol{\mu}\right)^T\Sigma^{-1}\left(\boldsymbol{x}-\boldsymbol{\mu}\right)}\right),
\end{align}
where $v=(2-d)/2$, and $K_v$ is the modified Bessel function of the second kind and of order $v$.
A key feature of the symmetric ML distribution is that its marginal distributions are Laplace distributions.
Sampling from an ML distribution is equivalent to sampling from a normal distribution with stochastic variance, where the variance is distributed according to the variance of the ML distribution multiplied by the square root of a random variable (r.v.) distributed according to an exponential distribution with scale 1~\cite{kotz2001laplace}.
Let $\boldsymbol{Y}$ be the r.v. from the symmetric ML distribution with mean $\boldsymbol{\mu}$ and variance $\Sigma$, $\boldsymbol{X}$ is the r.v. from the multivariate Gaussian distribution with mean $\boldsymbol{0}$ and variance $\Sigma$, and $\boldsymbol{Z}$ is the r.v. distributed according to an exponential distribution with scale~1.
Then:
\begin{equation}
    \boldsymbol{Y} = \sqrt{Z}\boldsymbol{X} + \boldsymbol{\mu}.
\end{equation}
This relationship can trivially be written as the following integral:
\begin{equation}
    \int_0^\infty e^{-z}\, \frac{1}{\sqrt{|2\pi\,z\,\Sigma|}} \, e^{-\frac{1}{2}\left(\boldsymbol{x}-\boldsymbol{\mu}\right)^T(z\,\Sigma)^{-1}\left(\boldsymbol{x}-\boldsymbol{\mu}\right)} \,dz,
\end{equation}
which in turn is the following CGMM:
\begin{equation}
\label{eq:MLCGMM}
    \int_0^\infty e^{-z}\,\mathcal{N}\left(\boldsymbol{x}; \boldsymbol{\mu}, z\,\Sigma\right) \,dz,
\end{equation}
where the p.d.f. $p_z(z)$ is $e^{-z}$. The equation shows that the variance of each mixand increases linearly with the parameter~$z$.

The ML as an infinite Gaussian mixture belongs to the class of the GSMs~\cite{boris2008scale}, defined as
\begin{equation}
    p(\boldsymbol{x}) = \int_0^\infty \mathcal{N}\left(\boldsymbol{x};\boldsymbol{\mu}+z\boldsymbol{\beta}, \Sigma/\kappa(z)\right) p_z(z)    \,dz,
\end{equation}
where $\beta$ is a shape parameter and $\kappa\left(\cdot\right)$ is a positive scale function. In addition to the ML distribution, several others are known to have representations as GSMs, such as the Cauchy distribution and Student's $t$ distribution. GSMs enjoy properties that make them more tractable than general CGMM. After even a linear time update though, the ML-CGMM is no more a GSM, but just a general CGMM. The previously mentioned filter by Huang et al.~\cite{huang2017robust} approximates the transitional prior of a Gaussian distribution with ML process noise as a GSM.
\section{Quadrature and Interpolation}
\label{sec:interpolation}
Quadrature allows one to compute an approximation to the p.d.f. of the CGMM in finite time. This computation is required whenever one wants to reduce the CGMM to a single Gaussian distribution. This is different from quadrature or cubature filters such as the cubature Kalman filter~(CKF)~\cite{cqf} or the Unscented Kalman Filter~(UKF)~\cite{julier1997unscented}, since here the quadrature is done over an independent parameter that describes a non-Gaussian distribution. At the same, by interpolation one can obtain an approximation to the original CGMM while only saving the value at a few nodes. This way any transformation, such as time update or measurement update, can be performed in a finite amount of time.
Interpolation is also useful to switch the number of nodes when performing different operations; for example, for astrodynamics problems the time update is generally more time consuming than the measurement update, and thus one may want to have fewer nodes for the time update and more nodes for the measurement update.

Quadrature allows to efficiently compute the integral \eqref{eq:CGMMdef}. As the integral for the ML-CGMM is indefinite, particular attention needs to be paid to the choice of the quadrature nodes. Gauss-Laguerre quadrature is used to numerically compute the integral
\begin{equation}
    \int_0^\infty e^{-z}\,f(z)\,dz \approx \sum_{i=1}^n w_i\,f(z_i).
\end{equation}
In this case, $f(z)=w(z)\,\mathcal{N}\left(\boldsymbol{x}; \boldsymbol{\mu}(z),\Sigma(z)\right)$.
The $n$ nodes $z_i$ for Gauss-Laguerre quadrature are the roots of the Laguerre polynomial $L_n(x)$:
\begin{equation}
    L_n(z) = \frac{1}{n!}\left(\frac{d}{dz}-1\right)^n z^n,
\end{equation}
for integer $n$, and the corresponding interpolation weights $w_i$ are computed as
\begin{equation}
\label{eq:GaussLaguerreWeights}
    w^i = \frac{z_i}{\left(n+1\right)^2\left[L_{n+1}\left(z_i\right)\right]^2}.
\end{equation}

It is possible to recover an approximation to the full distribution from just the values at a few nodes by interpolation. Spline interpolation is preferred here for simplicity. The interpolation nodes do not need to be the same as the Gauss-Laguerre quadrature nodes; however, one needs to choose the $n_i$ interpolation nodes $\left[z_1, \dots , z_{n_i}\right]$ such that any following evaluations of the interpolation do not lie outside of the interval $\left[z_1,z_{n_i}\right]$.
Spline interpolation can directly be used for the means of the mixands. The interpolation of the p.d.f. $p_z(z)$ can be done by interpolating its natural logarithm, so that positivity is ensured. The interpolated function needs then to be normalized such that its integral is equal to 1. The covariance can be interpolated in several ways. One way consists of taking the Cholesky decomposition, and interpolate it element-by-element. Another way would consist of, after taking the Cholesky decomposition, generating the $\sigma$-points as in~\cite{julier1997unscented}, and then interpolating those. In both cases positive semidefiniteness and symmetry are preserved, but some of the eigenvalues may still be zero. If the interpolation is done over the $\sigma$-points, then it can also be used to recover the means of the mixture mixands.
\section{The GIF with ML Process Noise}
Consider the nonlinear stochastic discrete-time system with non-additive process noise
\begin{align}
\label{eq:nonlinearsystemdef}
    \boldsymbol{x}_{k} &= \boldsymbol{f}_k\left(\boldsymbol{x}_{k-1},\boldsymbol{v}_{k-1}\right),\\
    \boldsymbol{y}_{k} &= \boldsymbol{h}_k\left(\boldsymbol{x}_{k}\right) + \boldsymbol{w}_{k},
\end{align}
where $\boldsymbol{x}_{k}$ is the state of the system at time $k$, $\boldsymbol{f}_k\left(\cdot\right)$ is a transition function, $\boldsymbol{y}_{k}$ is the measurement at time $k$, $\boldsymbol{h}\left(\cdot\right)$ is the measurement function, and $\boldsymbol{v}_{k-1}$ and $\boldsymbol{w}_{k}$ are random variables.
For the case where the random variables are Gaussian, this problem can be approximately solved by an Extended Kalman Filter (EKF) or a UKF, which perform, respectively, local and statistical linearization. In this paper we consider the case in which $\boldsymbol{v}_{k-1}$ are distributed according to an ML distribution. The case where  instead $\boldsymbol{w}_{k-1}$ follows an ML distribution is not treated here, but the solution method is very similar.
First, the number of interpolation nodes $n_t$ to use during propagation needs to be decided. Then, assuming the prior at time $k-1$ is Gaussian, the distribution is propagated for every node, either using the UT, like for a UKF, or by linearizing around the mean, like in the EKF:
\begin{align}
    \boldsymbol{x}_{k|k-1} &= f(\boldsymbol{x}_{k-1|k-1}),\\
    P^i_{k|k-1,t} &= F_k P_{k-1|k-1} F_k^T + \Gamma_k \left(z^i_{t}Q\right) \Gamma_k^T,
\end{align}
where the superscript $i$, together with the subscript $t$, means that the value is for the $i$\textsuperscript{th} time update node, $P^i_{k|k-1}$ is the transitional prior covariance at time $k$, $P_{k-1|k-1}$ is the prior covariance at time $k-1$, $F_k$ is the state transition matrix, $z_{i,t}$ is the value of $z$ at node $i$ for the time update, $Q_k$ is the covariance of the ML process noise, and $\Gamma_k$ is the process noise Jacobian. Note that the components' weights are not considered yet. When using EKFs and starting with a Gaussian distribution at time $k-1$, the computations of $f(\boldsymbol{x}_{k-1|k-1})$,  $F_k$, and $\Gamma_k$ are the same for any $i$, since they all take the same input $\boldsymbol{x}_{k-1|k-1}$. Those computations can thus be carried out just once, regardless of how many mixands are propagated, making the time update negligibly larger than that of a single EKF. In a similar fashion, if a bank of UKFs is used instead of a bank of EKFs, the different mixands share some of the $\sigma$-points, since the noise is uncorrelated from the state; specifically, only $2\,\text{dim}\left(\boldsymbol{v}\right)$ points need to be computed for every mixand other than the first one.
After propagation the time update nodes are switched to the measurement update nodes. The number of mixands $n_m$ for the measurement update is usually larger than $n_t$. The values at the new nodes can be found by interpolation, as discussed in Sec.~\ref{sec:interpolation}. For the EKF, the only variable to be interpolated is the covariance $P^i_{k|k-1}$:
\begin{equation}
    S_{L_{k|k-1}}(z) = S_{L_{k|k-1}}\left(z|L^1_{k|k-1,t},\ldots,L^{n_t}_{k|k-1,t}\right),
\end{equation}
where $L^i_{k|k-1}$ is the lower triangular Cholesky decomposition of $P^i_{k|k-1}$, and $S_y\left(z|M^1,\ldots,M^n\right)$ is a function interpolating the data matrices $M^1,\ldots,M^n$ at nodes $z^1,\ldots,z^n$, and evaluated at $z=z$. The transitional prior covariances at the measurement nodes are then computed:
\begin{equation}
\label{eq:covInterpmeasurement}
P^i_{k|k-1,m}=S_{L_{k|k-1}}(z^i_m)\left(S_{L_{k|k-1}}(z^i_m)\right)^T.    
\end{equation}

The measurement update for the bank of EKFs is then:
\begin{align}
    \boldsymbol{\Delta}\boldsymbol{y} & = \boldsymbol{y} - \boldsymbol{h}(\boldsymbol{x}_{k|k-1}),\\
    S^i_{k|k-1} & = H_k P^i_{k|k-1,m}H_k^T + R_k,\\
    K^i_k & = P^i_{k|k-1,m}H_k^T\left(S^i_{k|k-1}\right)^{-1},\\
    \boldsymbol{x}^i_{k|k,m} & = \boldsymbol{x}^i_{k|k-1} + K^i_k\boldsymbol{\Delta} \boldsymbol{y},\\
    P^i_{k|k,m} & = \left(I-K_k^iH_k\right)P^i_{k|k-1,m},\\
    l^i_{k|k,m} &= \frac{1}{\sqrt{|2\pi S^i_{k|k-1}|}}\,e^{-1/2\,\boldsymbol{\Delta} \boldsymbol{y}^T(S^i_{k|k-1})^{-1}\boldsymbol{\Delta} \boldsymbol{y}}
\end{align}
where the superscript $i$, together with the subscript $m$, means that the variable is for the $i$\textsuperscript{th} measurement update node (the subscript $m$ is avoided for variables that do not show up during time update or quadrature), $S^i_{k|k-1}$ is the innovation covariance, $K^i_k$ is the gain matrix, $\boldsymbol{y}$ is the actual measurement, and $l^i$ is the measurement likelihood. Note that, as in the time update, some computations are the same for all components: the expected measurement $\boldsymbol{h}(\boldsymbol{x}_{k|k-1})$ and the measurement Jacobian $H_k$.

Finally, quadrature is needed to obtain an actual approximation to the posterior.
To compute the posterior in finite time, it is represented as a GMM. Nonetheless, at any time, an approximation to the original CGMM can be recovered back. The $n_q$ quadrature nodes are interpolated from the measurement update nodes. As one can never interpolate outside of the data bounds, it should be made sure that all successive interpolation extrema are inside the previous ones: $\left[z_{t,1},z_{t,n_t}\right]\in\left[z_{m,1},z_{m,n_m}\right]\in\left[z_{q,1},z_{q,n_q}\right]$.
Interpolating functions are used again. The variables to interpolate are the lower triangular Cholesky decompositions $L^i_{k|k}$ of the posterior covariances $P^i_{k|k}$, the means of the posterior distributions $\boldsymbol{x}^i_{k|k}$, and the log-likelihoods $\log l^i$:
\begin{align}
    S_{L_{k|k}} (z) & = S_{L_{k|k}} \left( z|L^1_{k|k,m}, \ldots, L^{n_t}_{k | k , m} \right),\\
    \boldsymbol{s}_{x_{k|k}}(z) &= \boldsymbol{s}_{x_{k|k}}\left(z| \boldsymbol{x}^1_{k|k,m}, \ldots, \boldsymbol{x}^{n_t}_{k|k,m} \right),\\
    {s}_l(z) &= {s}_l\left(z|\log l^1_{m},\ldots,\log l^{n_t}_{m}\right),
\end{align}

The mean of each quadrature component is simply evaluated from the interpolation, and the covariance is computed in a similar fashion as \eqref{eq:covInterpmeasurement}. The relative weights are computed as follows:
\begin{equation}
    \hat{w}^i_{k|k,q} = w^i_q\,e^{s_l\left(z^i_q\right)},
\end{equation}
where $w^i_q$ is the quadrature weight of the $i$\textsuperscript{th} quadrature node computed as in \eqref{eq:GaussLaguerreWeights}. Finally, the weights are normalized:
\begin{equation}
    {w}^i_{k|k}=\frac{{\hat{w}}^i_{k|k}}{\sum_{j=1}^{n_q}{\hat{w}}^j_{k|k}}.
\end{equation}
The method can similarly be applied using UKFs instead of EKFs. In that case, $2\text{dim}(\boldsymbol{x})+1$ $\sigma$-points can be reused for all components after the first, since only the process noise changes between nodes.

\section{Results}
In the results section we first analyze a simple linear problem, and look at how the results differ depending on whether $\text{rank}\left(H\right) = \text{dim}\left(\boldsymbol{x}\right)$ or $\text{rank}\left(H\right) < \text{dim}\left(\boldsymbol{x}\right)$. Then, we look at how the filter behaves in a complex maneuvering target tracking problem in Earth orbit, with large mismatch between the expected maneuver and the actual maneuver.
\subsection{Linear Case}
\begin{figure}
\centering
\includegraphics[trim= 15mm 5mm 15mm 10mm, clip,width=\linewidth]{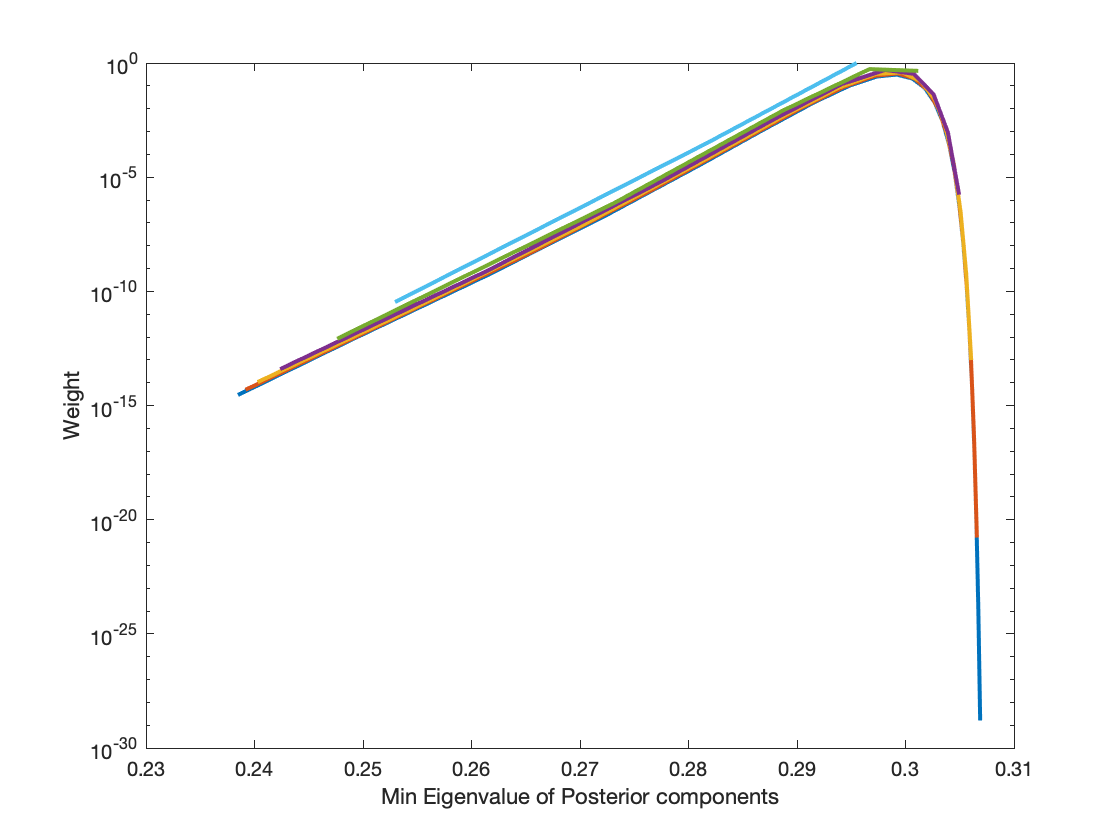}
  \centering
  \includegraphics[trim= 15mm 5mm 15mm 10mm, clip,width=\linewidth]{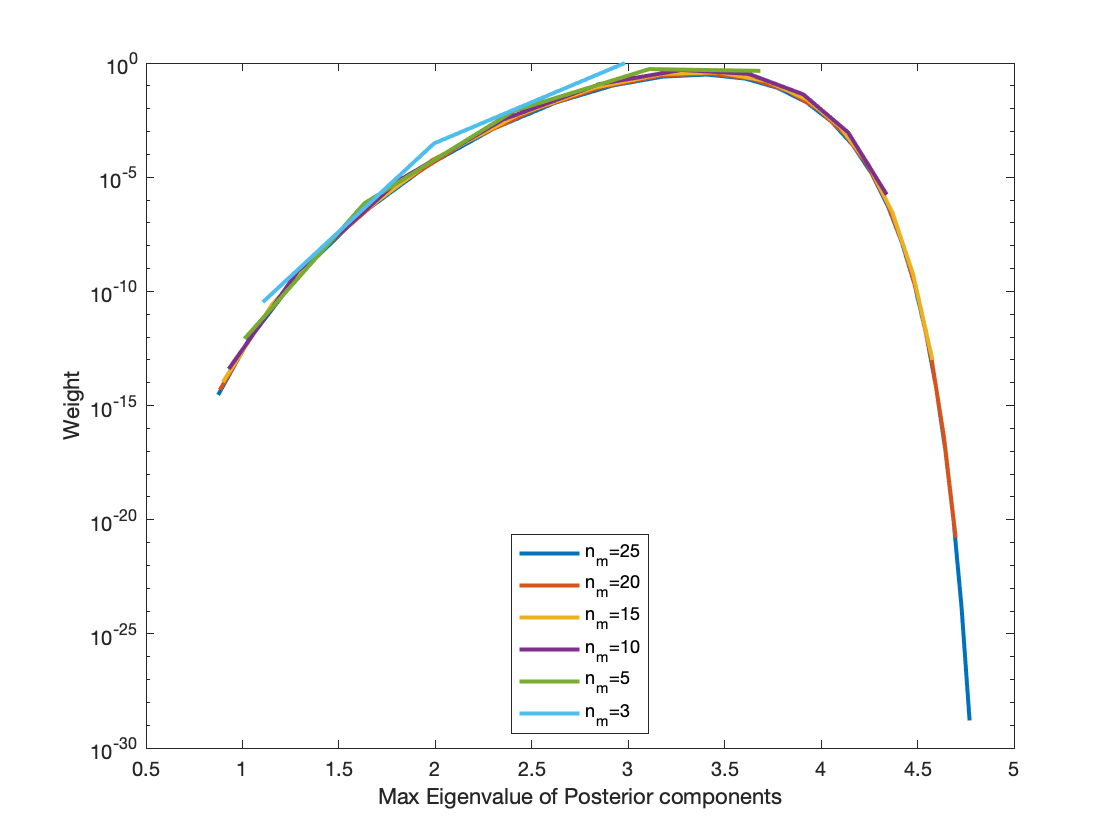}
\caption{Posterior weights of the mixands versus smallest (top) and largest (bottom) eigenvalues of their variance, for different values of $n_m$, when $\text{rank}\left(H\right) = \text{dim}\left(\boldsymbol{x}\right)$.}
\label{fig:hnonsing}
\end{figure}
The first application is a simple toy problem only aimed at demonstrating the behavior of the ML-GIF with an ML prior and a Gaussian measurement. No tracking is involved here.
Consider the following linear 3-dimensional problem:
\begin{align}
    \boldsymbol{x}_{k} = F_k \boldsymbol{x}_{k-1} + \Gamma_k\boldsymbol{v}_{k-1},\\
    \boldsymbol{y}_{k} = H_k \boldsymbol{x}_{k} + \boldsymbol{w}_{k},\\
\end{align}
with $F_k  =  \Gamma_k  = \mathbb{I}_{3\times 3}$, and
\begin{equation}
    \nonumber H_k = \begin{bmatrix} 1 & 0 & 1\\
    0 & 1 & 0\\
    0 & 1 & 1
    \end{bmatrix},
\end{equation}
 and the process noise $\boldsymbol{v}_{k-1}$ is distributed according to an ML with variance $Q_k=\mathbb{I}_{3\times 3}$ and mean $\boldsymbol{0}$, and the measurement noise $\boldsymbol{w}_{k}$ is Gaussian with variance $R_k=\mathbb{I}_{3\times 3}$ and mean $\boldsymbol{0}$.
At epoch $k-1$ the prior distribution for the state $\boldsymbol{x}_{k-1}$ is set to have mean $\boldsymbol{x}_{k-1|k-1}=\boldsymbol{0}$ and covariance $P_{k-1|k-1}=\mathbb{I}_{3\times 3}$. Here $\text{rank}\left(H\right) = \text{dim}\left(\boldsymbol{x}\right)$, and thus we expect the posterior to be sub-Gaussian. Assume now that the measurement $\boldsymbol{y}_k=\left[0,-15,-6\right]$ is obtained. 

Fig.~\ref{fig:hnonsing} shows the posterior weight of each mixand versus the minimum and maximum eigenvalue of the posterior covariance, for several choices of $n_m$. For increasing magnitude of the eigenvalue, both plots reach what seems to be a vertical asymptote whenever $n_m$ is set to be larger than 5. For this specific problem $n_m=10$ seems to be large enough, in the sense that all additional mixands for $n_m>10$ have very small weights. However, in case the deviation were even larger, more nodes may be necessary: the larger the number of nodes, the better a large deviation can be tracked.

\begin{figure}
\centering
\includegraphics[trim= 15mm 5mm 15mm 10mm, clip,width=\linewidth]{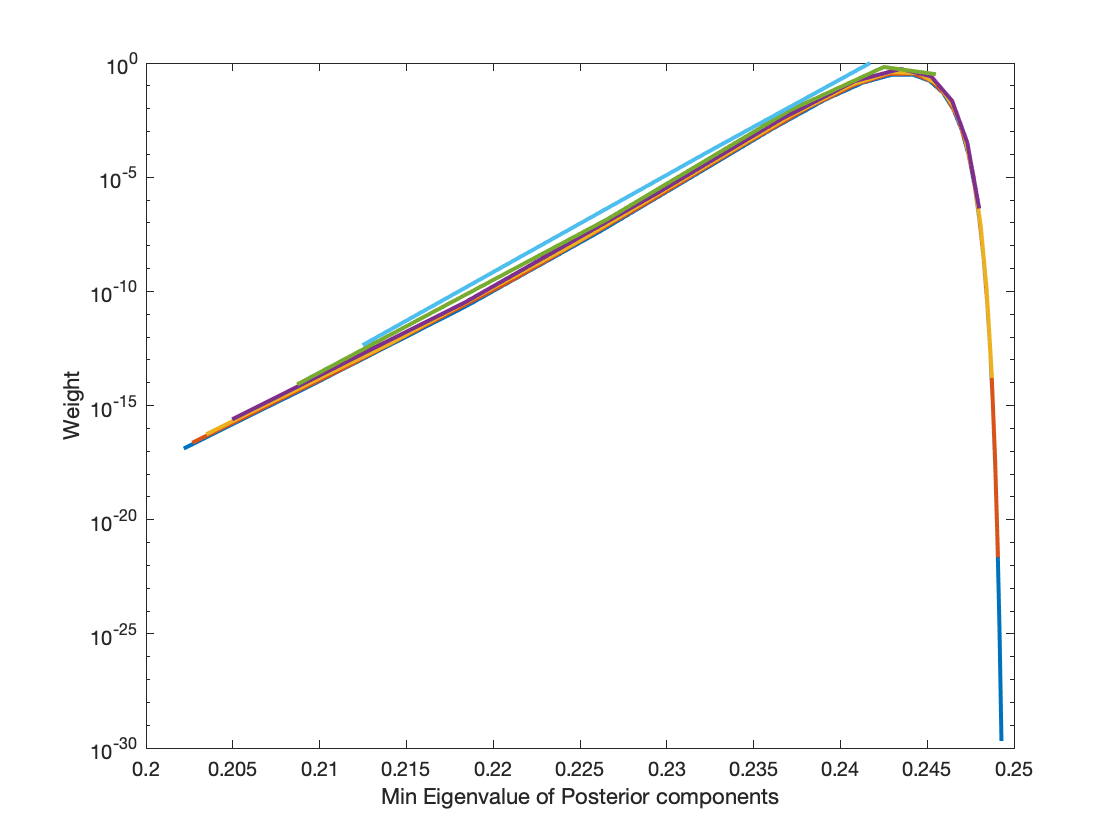}
  \centering
  \includegraphics[trim= 15mm 5mm 15mm 10mm, clip,width=\linewidth]{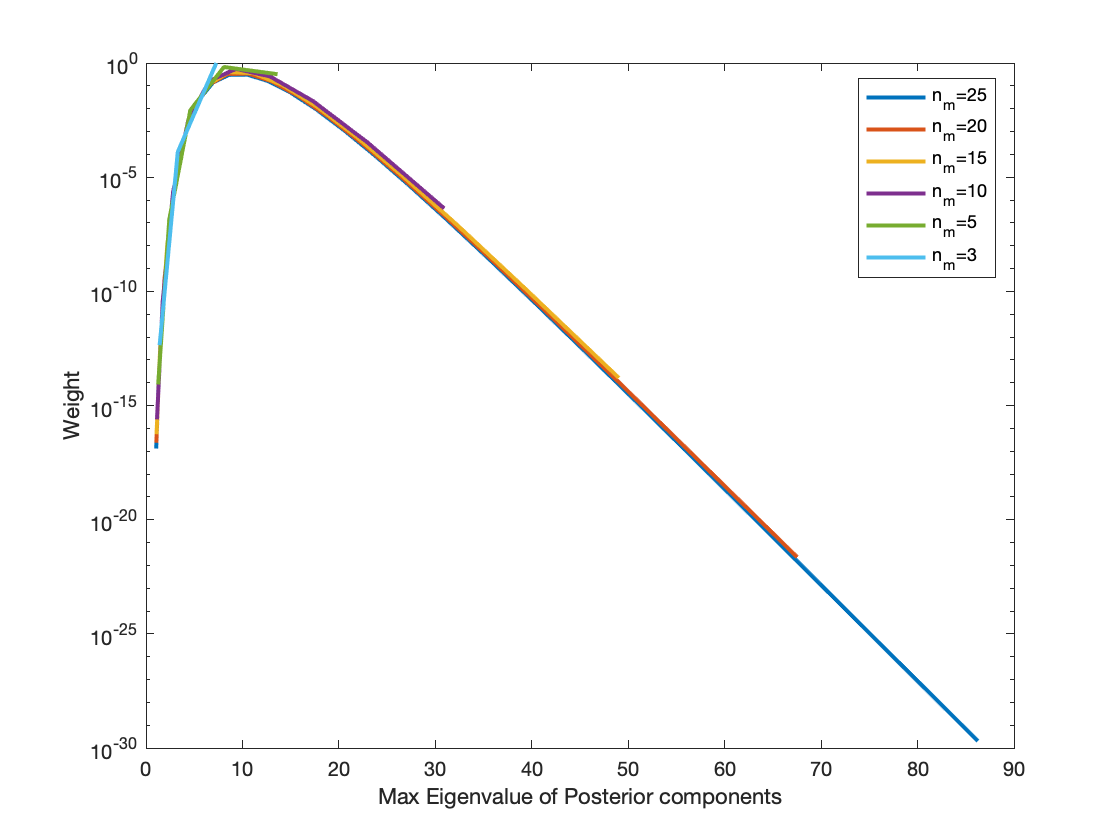}
\caption{Posterior weights of the mixands versus smallest (top) and largest (bottom) eigenvalues of their variance, for different values of $n_m$, when $\text{rank}\left(H\right) < \text{dim}\left(\boldsymbol{x}\right)$.}
\label{fig:hsing}
\end{figure}
Let us consider now the case where the rank of the $H_k$ matrix is smaller than the dimensionality of $\boldsymbol{x}$:
\begin{equation}
\nonumber H_k = \begin{bmatrix} 1 & 0 & 1\\
    0 & -1 & 0\\
    1 & 0 & 1
    \end{bmatrix}.
\end{equation}
Fig.~\ref{fig:hsing} shows plots for the same variables as the previous figure, but for the latter case. Here the largest eigenvalue of the variance increases linearly with the logarithm of the weight. As per \eqref{eq:MLCGMM}, if the variance increases linearly with the logarithm of $p_z(z)$, then the distribution is, along at least one dimension, an ML. If the linear relation only occurs for some values of $z$ larger than a certain threshold, as is the case for the largest eigenvalues, then one can state that the tail is that of an ML distribution. If this is true for at least one eigenvalue, it means that there is a decomposition such that the distribution is heavy-tailed along at least one dimension. Hence, the plot shows that the posterior is still heavy-tailed along at least one of its dimensions. In contrast, the smallest eigenvalue still reaches what seems to be an asymptote, showing that at least one of the dimensions has sub-Gaussian tails, as expected.

\subsection{Low-Thrust Maneuvering Spacecraft Tracking with Sparse Observations}
Low-thrust maneuvering spacecraft tracking is more challenging than traditional maneuvering target tracking problems because it involves sparse observations and continuous thrust, which keep the uncertainty large for long periods of time~\cite{kelecy2010detection}.
In this subsection, we analyze the results obtained for the tracking of a low-thrust maneuvering spacecraft that is spiraling out with constant in-track thrust. After the scenario description, the results are analyzed for the case where the GIF's nodes are kept constant between time update, measurement update, and quadrature. Then, different combinations of time update nodes and measurement update nodes are tested. In all cases, a bank of UKFs is used, and the integral of mean and covariance is computed after every measurement: the posterior state is always reduced to a Gaussian distribution. The propagation is performed with 19~$\sigma$-points, because the state has 6 dimensions and the process noise has 3~dimensions. After the propagation is carried out for the first mixand, all other mixands only need 6~$\sigma$-points to be propagated, as the other 13 are shared among all mixands, since they do not include the process noise. Hence, propagation time for 10~nodes only takes about 4~times the computational resources of a single UKF.

The only forces in play in this scenario are the central gravity, perturbation due to $J_2$, and thrust:
\begin{align}
    \boldsymbol{a} = -\frac{\mu}{r^3}\boldsymbol{r} +\boldsymbol{a}_{J_2}+\boldsymbol{T},
\end{align}
where $\mu$ is the gravitational parameter of Earth, $\boldsymbol{r}$ is the $\left[x,y,z\right]$ position of the spacecraft, $\boldsymbol{T}$ is the thrust, and $\boldsymbol{a}_{J_2}$ is the acceleration due to $J_2$:
\begin{align}
    a_{J_2,x} &= -\frac{3}{2}\mu\, J_2 \,\frac{R_e^2}{r^5}\left(1-5\,\frac{z^2}{r^2}\right)\,x,\\
    a_{J_2,y} &= -\frac{3}{2}\mu\, J_2 \,\frac{R_e^2}{r^5}\left(1-5\,\frac{z^2}{r^2}\right)\,y,\\
    a_{J_2,z} &= -\frac{3}{2}\mu\, J_2 \,\frac{R_e^2}{r^5}\left(3-5\,\frac{z^2}{r^2}\right)\,z,
\end{align}
where $R_e$ is the Earth's Equatorial radius, and $J_2$ is the coefficient of degree 2 and order 0 of the spherical harmonics expansion describing Earth's gravity field. The thrust is treated by the filter as the random variable $\boldsymbol{v}_{k-1}$ from \eqref{eq:nonlinearsystemdef}, distributed as an ML. The initial conditions are distributed according to
\begin{align}
   \nonumber \boldsymbol{x}_{0|0} &= \begin{bmatrix}
        0~\text{km} & 7,000.000~\text{km} & 0~\text{km}
    \end{bmatrix},\\
  \nonumber  \dot{\boldsymbol{x}}_{0|0} &= \begin{bmatrix} 5,335.865~\text{m/s} & 0~\text{m/s} & 5,335.865~\text{m/s}
    \end{bmatrix},\\
  \nonumber  P_{0|0} &= \begin{bmatrix}
100~\mathbb{I}_{3\times 3}~\text{m} & 0 \\
0 & 0.1~\mathbb{I}_{3\times 3}~\text{m/s} \end{bmatrix}^2.
\end{align}
One radar measurement is performed every 10,000~s, which is a little less than twice the initial orbital period. To keep the scenario simple, the measurement is simulated as coming from the center of the Earth, and consists of range $\rho$, range-rate $\dot{\rho}$, right ascension $\alpha$, and declination $\delta$. The measurement error variance is
\begin{equation}
  \nonumber  R=\left(\text{diag}\begin{bmatrix}
3~\text{m} &
   0.03~\text{m}/\text{s} &
   0.015~\text{deg} &
   0.015~\text{deg}        
    \end{bmatrix}\right)^2.
\end{equation}
The measurement model provides direct information on the position with a standard deviation of approximately 2.5~km, whereas only one dimension of the velocity is observed at a time. This makes the problem unobservable without a prior.
The spacecraft accelerates with continuous thrust of 300~$\mu$m/s\textsuperscript{2} in the along-track direction, spiraling out. The magnitude and direction of the thrust are unknown to the filter. The filter assumes that the acceleration is constant between two successive observations, but that it can change after any measurement; moreover, it has no memory of the previous thrust profile, to maximize responsiveness. In this scenario the filter assumes that the standard deviation of the thrust is 10~$\mu$m/s\textsuperscript{2}, 30~times smaller than the actual one, to stress the capability of the ML-GIF when the target's acceleration magnitude is unknown. All computations were performed in Matlab, with a single thread of a 2.8 GHz Quad-Core Intel Core i7 processor.
\subsubsection{Constant Nodes}
\begin{figure}
\centering
\includegraphics[trim= 15mm 15mm 15mm 15mm, clip,width=\linewidth]{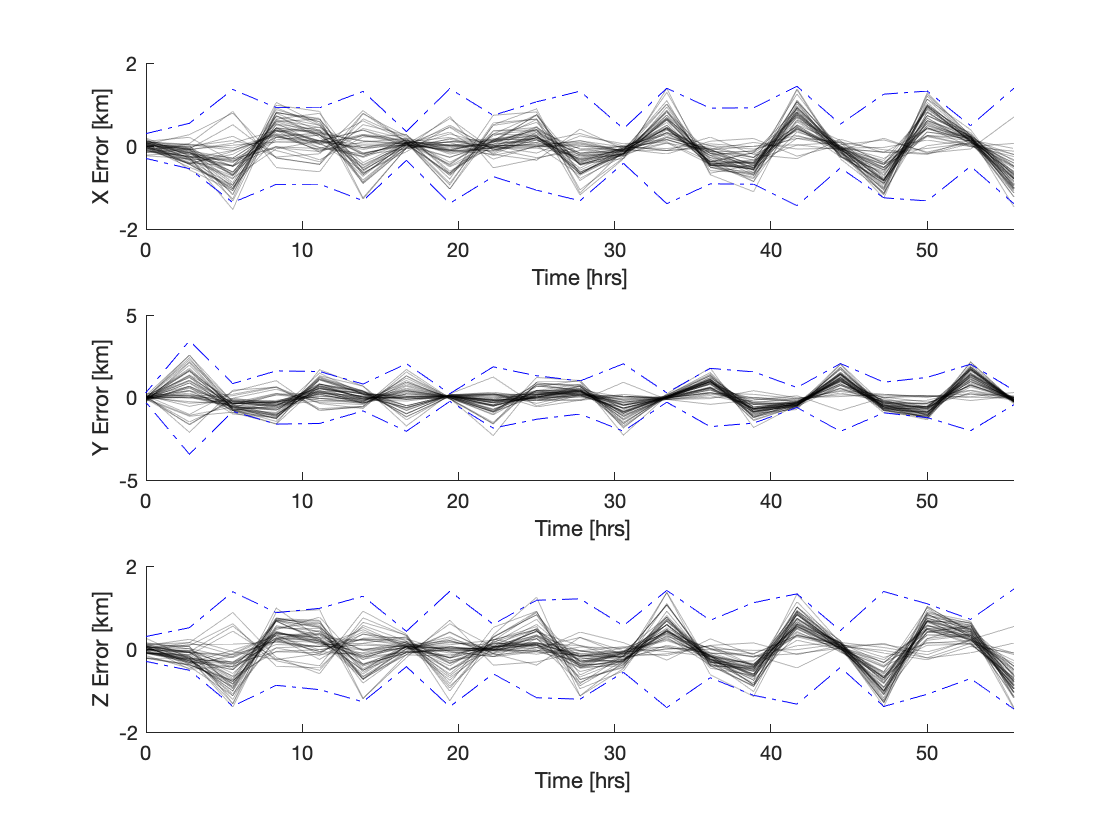}
  \centering
  \includegraphics[trim= 15mm 15mm 15mm 15mm, clip,width=\linewidth]{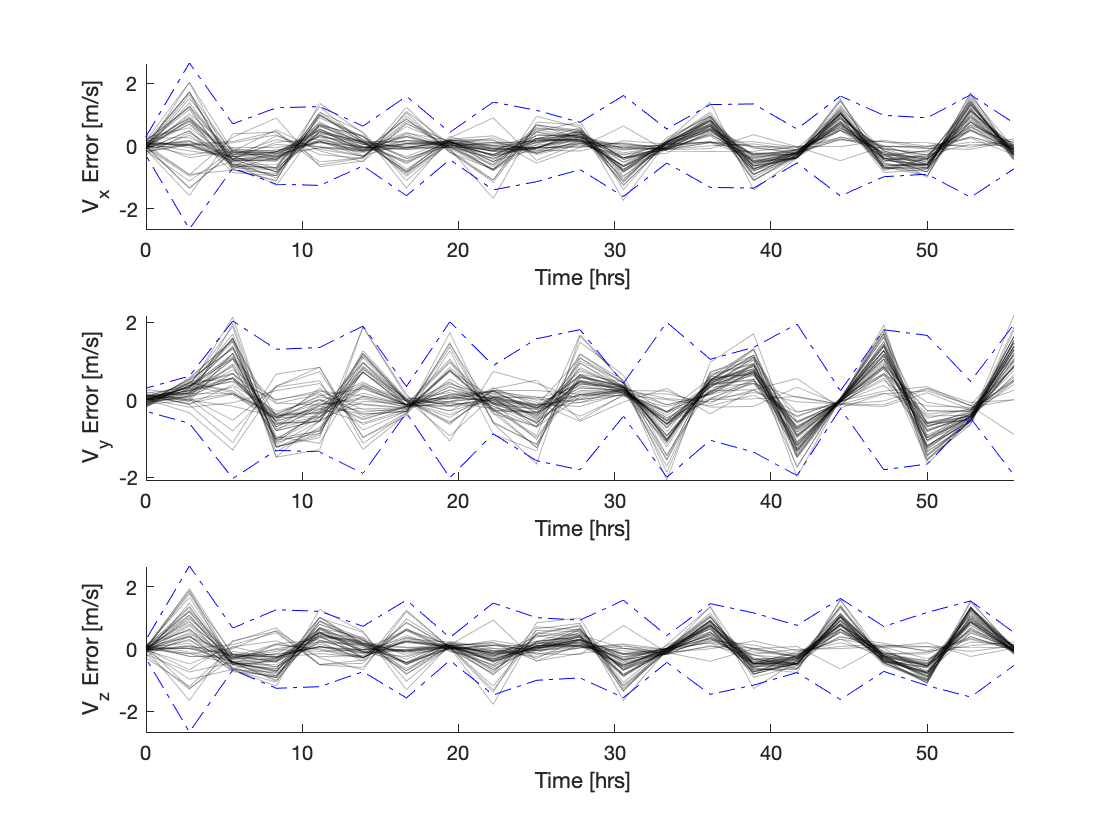}
\caption{Position~(top) and velocity~(bottom) errors for the orbital case with in-track thrust equal to 300~$\mu$m/s\textsuperscript{2}, using the ML-GIF. Dark blue is the average predicted 3$\sigma$ uncertainty.}
\label{fig:posvel25}
\end{figure}
\begin{figure}
\centering
  \includegraphics[trim= 15mm 5mm 15mm 10mm, clip,width=\linewidth]{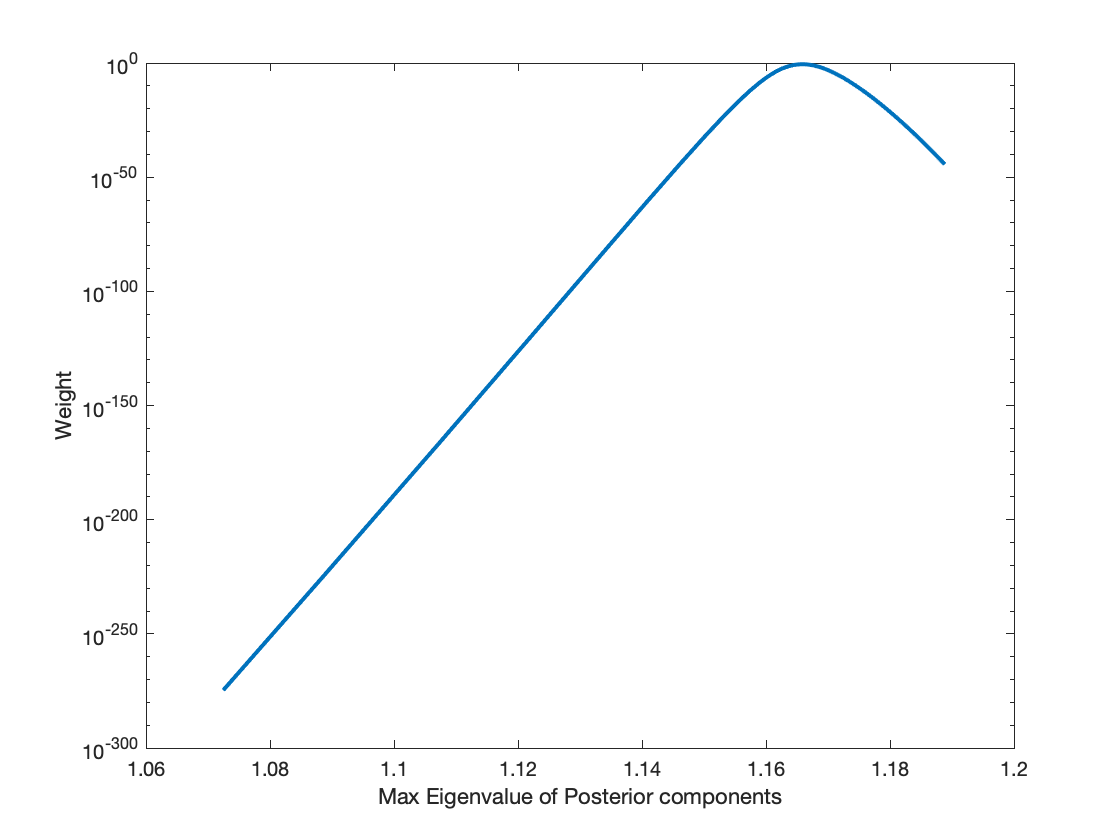}
\caption{Maximum eigenvalue of the mixands' posterior covariance versus their weights. The plot is for the first observation of the first run of the Monte Carlo trials. This specific run has been performed with 40~nodes to better show the trend.}
\label{fig:eig20orbit}
\end{figure}
\begin{figure}
\centering
  \includegraphics[trim= 15mm 15mm 15mm 15mm, clip,width=\linewidth]{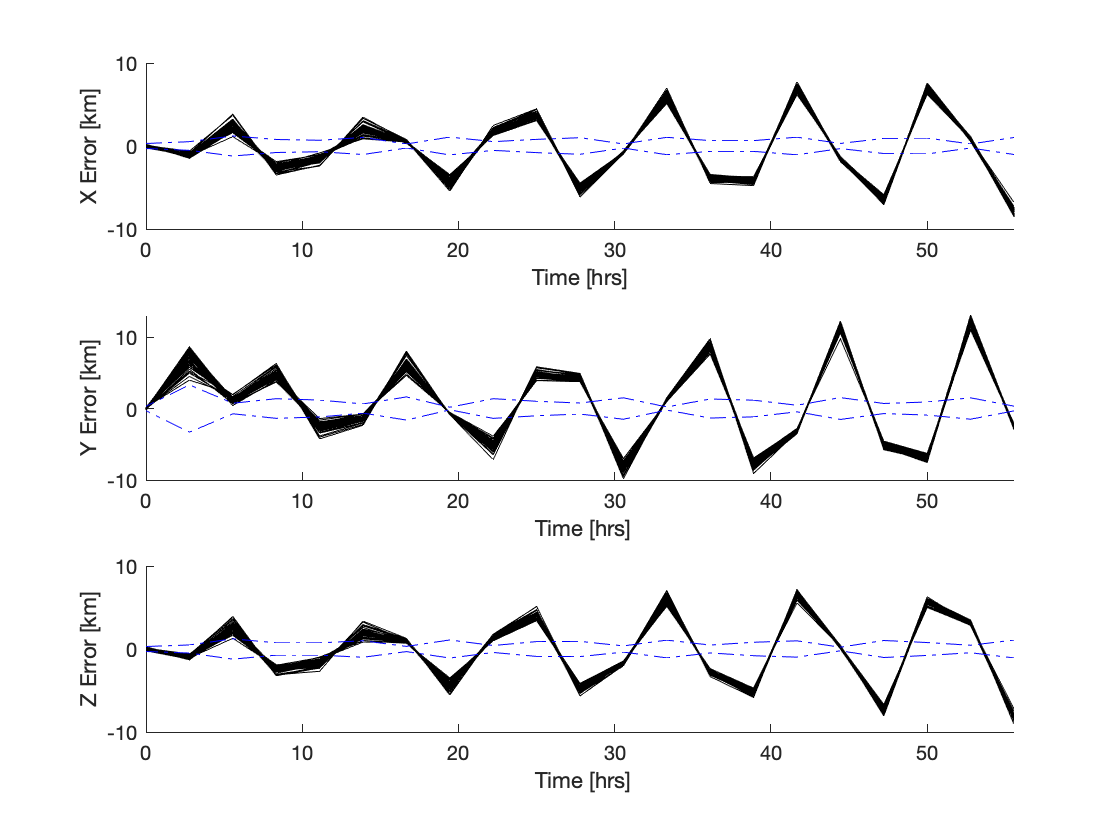}
\caption{Position errors for the orbital case with in-track thrust equal to 300~$\mu$m/s\textsuperscript{2}, using a UKF. Dark blue is the average predicted 3$\sigma$ uncertainty.}
\label{fig:posGauss25}
\end{figure}

For this case the nodes used for time update, measurement update, and quadrature are the roots of the Laguerre polynomial of order~10. Using a lower number of nodes leads to situations where the highest weighed quadrature component is also the one with the largest initial variance, causing the filter to miss relevant portions of the distributions. The computational time over the 50~runs is 1,532~s.
Fig.~\ref{fig:posvel25} shows the error in position and velocity obtained over 50~Monte Carlo trials, together with average 3$\sigma$ filter uncertainty. The error shows a bias, different at every measurement epoch, caused by the fact that the constant thrust introduces a systematic error in the model. About 1.72\% of the measurements fall outside of the 3$\sigma$ predicted variance. As the posterior resembles a Laplace distribution along at least one of the dimensions, as implied by Fig.~\ref{fig:eig20orbit}, around 1.5\% of estimates are expected to be outside the 3$\sigma$ bounds. While the frequency is slightly larger, this is acceptable considering the fact that a large systematic error is involved. Moreover, note that a majority of large deviations occur during the first few estimates, when the filter is still adjusting to the initial variance. Even though from the plot it looks like the uncertainty increases in the beginning, the determinant of the variance actually decreases, because correlation between the states is introduced by the measurements and the dynamics. This is a known occurrence for orbital problems starting with diagonal covariance matrices~\cite{woodburn}.
The position RMSE over all runs and epochs is 1,037~m, and the velocity RMSE is 1.096~m/s. As a reference, for this problem, after just the first observation the position and velocity of the accelerating satellite differ from those of a ballistic satellite by, respectively, 44~km and 48~m/s.
To compare, Fig.~\ref{fig:posGauss25} shows the performance of a single UKF with same process noise variance as the ML-GIF. The RMSE is 8,625~m in position and 9.059~m/s in velocity, and 99.4\% of the state estimates fall outside of the 3$\sigma$~bounds. From the plot, one can clearly deduce that the Gaussian filter is diverging. The computational time required by the single UKF is 377~s. 
\subsubsection{Interpolated Nodes}
The same problem is now solved by interpolating the nodes between time and measurement update. For this case, the time update nodes differ from the measurement update nodes, but the measurement update nodes are chosen to be the same as the final quadrature nodes. Measurement update is not computationally demanding for this problem, and therefore there is no need to change nodes between measurement and quadrature. The first and last propagation nodes are always the same as the first and last chosen update nodes:
$z_{t,1}=z_{m,1}$, and $z_{t,{n_t}}=z_{m,{n_m}}$. The time update nodes in-between are chosen such that they are linear in a quadratic scale. Note that, for $n_t=2$ and $n_t=3$, spline interpolation is not possible, and linear and quadratic interpolations are used instead, respectively. The measurement update nodes are Gauss-Laguerre quadrature nodes, so that quadrature can directly be operated over the computed mixands. 

No plots are shown for these cases, because the results all look qualitatively very similar to the previous case. Table~\ref{table:interpolated_results} summarizes RMSE and computational time for every analyzed combination of $n_m=n_q$ and $n_t$. All combinations are evaluated over the same 50~Monte Carlo trials.
\begin{table}[htbp]
\caption{Performance over 50~Monte Carlo trials for ML-GIF with interpolation}
\label{table:interpolated_results}
\begin{center}
\begin{tabular}{|c|c|c|c|c|c|}
\hline
$n_t$ & $n_m$ & Pos. RMSE & Vel. RMSE & \% out 3$\sigma^{\mathrm{a}}$ & Comp. Time$^{\mathrm{b}}$\\
\hline
2 & 10 & 1,011~m & 1.068~m/s & 2.28 & 509~s \\
3 & 10 & 989~m & 1.046~m/s & 1.88 & 651~s \\
5 & 10 & 1,004~m & 1.062~m/s & 2.02 & 896~s \\
2 & 15 & 812~m & 0.862~m/s & 0.72 & 514~s \\
3 & 15 & 937~m & 0.992~m/s & 1.37 & 651~s \\
5 & 15 & 1,008~m & 1.065~m/s & 2.03 & 894~s \\
2 & 25 & 730~m & 0.774~m/s & 0.35 & 514~s \\
3 & 25 & 900~m & 0.955~m/s & 1.11 & 641~s \\
5 & 25 & 1,003~m & 1.061~m/s & 2.00 & 867~s \\
\hline
\multicolumn{6}{l}{$^{\mathrm{a}}$For a Laplace distribution, about 1.5\% of samples are outside 3$\sigma$.}\\
\multicolumn{6}{l}{$^{\mathrm{b}}$Total computational time for all 50 runs.}
\end{tabular}
\label{tab1}
\end{center}
\end{table}
The error introduced by the interpolation causes a difference in performance between the filters. Since the cases with 2 and 3 time update nodes use a different interpolation technique, namely linear and quadratic, instead of spline, it is impossible to conclude whether the difference in performance is caused by the different interpolation techniques or by the number of nodes. For same number of time update nodes, adding measurement nodes improves both accuracy and statistical consistency. Such improvement is smaller when going from 15~to 25~measurement nodes, likely because the acceleration of 30~standard deviations is captured well enough by 15~nodes. As expected, the main driver of the computational cost is the number of propagation nodes. The ML-GIF with $n_t=2$ and $n_m=n_q=25$ takes 1.4 times the computational resources of a single UKF, and performs better than the ML-GIF without interpolation with $n_t=n_m=n_q=10$, at little more than one third the computational cost. 
\section{Conclusions}
This paper introduces the GIF, the limit for the GSF when the number of mixands tends to infinity. The GIF is computed numerically by building on the framework of a GSF with quadrature and interpolation. Differently from a normal GSF, an approximation to the corresponding continuous mixture can always be obtained by interpolation. The interpolation can be used to reduce or increase  the number of discretization nodes, or to sample from the continuous distribution. While the GIF can be used for a variety of applications, this paper demonstrates the case in which an ML distribution is described as a CGMM, and used to represent the process noise of a maneuvering target. The resulting filter is able to discern whether the posterior distribution is heavy-tailed or not. The filter is successful in a simulated scenario consisting of a tracking problem with sparse observations where a satellite maneuvers with an acceleration that is 30~times the expected standard deviation. A Gaussian filter with same process noise variance diverges. The UKF-ML-GIF requires less than 1.5~times the computational cost of a UKF.

\FloatBarrier

\end{document}